\documentstyle[11pt,aaspp4]{article} \received{ ------- } \accepted{ 
------- }
\begin{document}

\title{Ohmic Decay of Magnetic Fields due to non-spherical accretion\\ in 
the
Crusts of Neutron Stars}

\author{ Mikael Sahrling} 
\affil{240 Cahuenga Dr, Oxnard, CA 93035}
\affil{micke@concentric.net} 
\begin{abstract} We 
consider
magnetic field evolution of neutron stars during polar-cap accretion. The size of the
polar cap increases as the field decays, and is set by the last open field line 
before the
accretion disk. Below the polar cap
we find the temperature to be so high that 
electron-phonon scattering
dominates the conductivity. Outside the polar cap region, the temperature 
is such
the conductivity is dominated by temperature independent impurity 
scattering
which can be a few orders of magnitude larger than the electron-phonon
conductivity. The time-scale for field decay is therefore initially given 
by
impurity scattering dominated conductivity. When the field strength has 
been
reduced to $\sim 10^8 ~{\rm gauss}$ the accretion is spherical and the 
time scale
for field decay is given by the smaller electron-phonon scattering 
conductivity. The field strength is now reduced rapidly compared to
before and this could be a reason for there being no pulsars known
with field strengths below $10^8~{\rm gauss}$.
We also investigate the evolution of multipoles at the neutron star 
surface. We
find that contribution from higher-order multipoles are at most 30 \% to 
that of
the dipole mode.

\end{abstract} 
\section{Introduction}
Since the discovery of pulsars there has been much discussion of
observational evidence 
for decay
of magnetic fields in neutron stars, as well as much theoretical work.  
As the
reviews by Lamb (1991), Chanmugam (1992), and Phinney \& Kulkarni (1994)
indicate, there is at present no consensus on the question of whether or 
not
magnetic fields in isolated neutron stars can decay significantly.  The general 
view has
been that the electrical conductivity of matter in the cores of neutron 
stars is
so high that the characteristic decay time for fields generated by 
electrical
currents in the core is greater than the age of the Universe. 
Recently Pethick \& Sahrling (1995) showed that even if the conductivity in the 
 core was small the shortest possible decay time is some two orders of
 magnitude longer than the decay time for configurations where the magnetic field
 is confined to the crust. An incorporation of general relativistic effects,
 Sengupta (1997), further reduces the decay rate. Still,
millisecond pulsars have typical field strengths in the range $10^8~ -~
10^{10}~{\rm gauss}$ compared to isolated radio pulsars which have typically
field strengths around $10^{12}~{\rm gauss}$ and this indicates
that during the spin-up phase the accretion process is reducing the
field strength somehow.

Millisecond pulsars are generally found in binary systems where the
companion star is a white dwarf with a mass less than a solar mass, $M_{\sun}$.
This system is called a Low-Mass-Binary Pulsar (LMBP) referring to the
mass of the companion star. 
The progenitor to this system is thought to be the Low-Mass-Xray Binaries 
(LMXB) where
a neutron star is accreting matter from a companion having a mass less than
about 2 $M_{\sun}$. The accreting matter is spinning up the neutron star to
millisecond periods. For details concerning this process see reviews by
Phinney \& Kulkarni (1994) and Bhattacharya \& van den Heuvel (1991) 
among others.
The evolution of the binary system after the neutron star is formed either
by a supernovae or tidal capture, is 
assumed to occur on at least two time scales when the companion star
is in radiative equilibrium. At first the companion star
is evolving on a nuclear time scale slowly filling its Roche lobe.
When it has been filled up the matter overflows and falls onto the companion
neutron star on a thermal time scale $\tau_{th}=G M^2/R L = 5\times10^7~
(M_{\sun}/M)^2 ~{\rm yrs}$, see Bhattacharya \& van den Heuvel (1991) for
details. 
If LMXB's are the only progenitors to LMBP's the lifetime of the LMXB,
or the accretion phase of the binary system, must be of order $10^7~{\rm
yrs}$. However, by using the amount of mass needed to be accreted to spin
up the neutron star to a spin period $P_i$
one finds a time scale that ranges from $10^8-10^{10}\times (P_i/2~{\rm ms}
)^{-4/3}~{\rm yrs}$. This discrepancy suggests that there might other 
progenitors to the LMBP's.
For details
see the review by Phinney and Kulkarni (1994).
 We will in this paper assume the accretion
phase of progenitors to millisecond pulsars to last between roughly $10^7-10^9~{\rm
 yrs}$.

Matter accreting onto a neutron star is expected to be disrupted by the 
magnetic
field at some radius, $r_A$, sometimes called the Alfv\'en radius. Beyond 
this
bare description there is no generally accepted view on how or where the 
matter 
attaches to the
field lines and flows to the neutron star's surface, or on the 
interaction of the
field with the matter outside $r_A$, despite a large
number of papers on the subject, see King (1995). In the case of disk 
accretion there 
are two main 
approaches to the problem. In
one (Ghosh \& Lamb 1978, 1979a,b; Kaburaki 1986 and Wang 1987) the field 
is assumed to
thread a large fraction of the disk because of Kelvin-Helmholtz 
instabilities.
 The other approach (Aly 1980; Anzer \& B\"orner 1980, 1983;
Scharlemann 1978) assumes the disk is a perfect conductor, completely 
excluding
the field. In both approaches the matter is often assumed to leave the 
disk in i
a narrow transition zone at the inner edge (near $r_A$), thereafter 
flowing
along field lines to the neutron star.

Most studies concerned with the magnetic field evolution {\it inside} the 
neutron star 
have assumed matter to accrete onto the surface spherically, e.g. 
Fujimoto et al. (1984), Miralda-Escud\'e et al. (1990),
hereafter Mir90, Urpin \& Geppert (1995), and Konar \& Bhattacharya 
(1997). In this paper I consider 
non-spherical
accretion where matter is assumed to flow onto the polar caps in a 
column. 
 The cap
is consequently heated up and the conductivity in the crust below the 
polar cap
we estimate to be much smaller than the conductivity outside the 
accretion column.
The magnetic evolution in this scenario can be divided into two stages 
where in stage
I the global decay rate is controlled by the conductivity outside
the accretion column, $\tau_B\sim 10^{8}-10^{10}~{\rm yrs}$. When the 
field has 
reached a value of about $10^8
{\rm gauss}$ the accretion is spherical and the evolution enters the 
second stage. 
Here, the whole crust is being heated up and the conductivity is 
dominated by
electron-phonon scattering. In this stage the magnetic field decay time 
is a few
orders of magnitude shorter than in stage I and compared to the earlier 
evolution the
field is dissipating rapidly. I argue that this could account for the 
fact that no
binary pulsars have been found with a magnetic field less than $10^8~{\rm 
gauss}$.
A similar effect where the accreted flow is pushing the field lines
has already been noticed by Romani (1993). 

The paper is organized as
follows: section 2 contains a brief description of the model and the 
basic
equations. In section 3 we estimate the length scale for temperature 
change at
the accretion column-normal crust boundary to be less than a crust 
thickness. The time
scale for the temperature to reach a stationary state is also shown to be 
much
smaller than the magnetic field decay time scale. Therefore, we do not 
solve the
energy equation explicitly but assume the conductivity as a function of 
angle to
be close to a top-hat function at all times. Section 4 presents the 
numerical
results for some initial depths of the magnetic 
field.

\section{Magnetic field evolution}

The purpose of the present  paper is to investigate the effect of 
asymmetric accretion on
the magnetic evolution in the {\it crust} of a neutron star. We therefore 
ignore
the complex interaction of matter and magnetic fields in the 
magnetosphere  and
for simplicity assume there is vacuum outside the star. The basic model 
we are
considering is an axi-symmetric spherical shell of outer radius $R_2$ and 
inner
radius $R_1$. Below $R_1$ there is a superconductor and outside the shell 
is
vacuum. I assume the equation of state in the shell to be given by Baym 
et al.
(1971). 
Furthermore, the magnetic field is assumed to vanish within the 
London
penetration depth of the superconductor.

The magnetic field, $\bf B$, obeys the usual equation
\begin{equation} {\partial{\bf B} \over\partial t} =
-{c^2\over{4\pi}}\nabla\times\left( {1\over\sigma}\nabla\times{\bf 
B}\right),
\end{equation} 
where we only include ohmic dissipation.
If one assumes the field to be initially poloidal, no 
torroidal field
will appear and one can write the magnetic field in terms of the vector
potential, ${\bf A} = A_\varphi {\bf e}_\varphi$ in spherical 
coordinates. One
finds the following equation for $A_\varphi$:
\begin{equation}\label{eq:Psi1} 
{\partial^2 A_\varphi\over{\partial 
r^2}}+{2\over
r}{\partial A_\varphi \over\partial r} + {1\over
{r^2}}{\partial\over\partial\theta} {1\over\sin\theta}{\partial A_\varphi
\sin\theta\over{\partial\theta}} = {4\pi\sigma(r,\theta)\over c^2} 
{\partial
A_\varphi\over { \partial t}}. 
\end{equation}

A detailed discussion of the boundary conditions used in this model can 
be found
in Sahrling (1996). As a summary, the boundary conditions to be imposed 
are
that, at the inner edge of the crust the field should be zero, no field 
diffuse
into the core, and at the outer boundary of the crust the poloidal field 
should
match onto the vacuum solution.

To estimate conductivities we shall use the calculations of the 
electrical
conductivity of matter in the crust of neutron star by Urpin \& Yakovlev 
(1980)
and Raikh \& Yakovlev (1982). To my knowledge, there exist no detailed 
studies of
phonons in the inner crust of neutron stars, where nuclei are immersed in 
a
neutron liquid.  However, one would expect the phonon spectrum to be 
qualitatively
similar to that for a lattice of ions, provided one replaces the mass of 
an ion
by the total mass of all nucleons in a cell of matter containing one 
nucleus. At high
temperatures, the dominant source of electrical resistivity is 
scattering of
electrons by phonons. One gets from Urpin \& Yakovlev (1980) the
corresponding conductivity  
$\sigma_{ph}\approx
1.5\times 10^{20}~{\rho_6/\mu_e} (1+(\rho_6/\mu_e)^{2/3})^{-1/2} 
1/T_8~{\rm
s^{-1}}$, where $\rho_6$ is the density in units of $10^6~{\rm g~ 
cm^{-3}}$, $\mu_e$ is the mean molecular weight per electron, $x$ is the
proton fraction of matter, and
$T_8$ is the temperature in units of $10^8~{\rm K}$. For temperatures 
less than
the Debye temperature, $T_D \approx 3.4 \times 10^9 \rho_{14}^{1/2} 
(x/0.1)K$,
electron scattering is due chiefly to Umklapp processes.  The 
corresponding
conductivity is $\sigma_{Um} \approx 5.5 \times 10^{23}~ \rho_{14}^{7/6}
~(x/0.1)^{5/3}T_9^{-2}~ {\rm s^{-1}}$ .  Below a temperature $T_U \approx 
2.2
\times 10^8 \rho_{14}^{1/2} (Z/60)^{1/3}(x/0.1)$ K, Umklapp processes are 
frozen
out, and the conductivity rises very sharply, since only normal processes 
contribute, 
and the electrical conductivity
due to phonon scattering is  $\sigma_{ph,N}\approx 2.1 \times
10^{28}\rho_{14}^{8/3} (x/0.1)^{14/3}T_9^{-5}$s$^{-1}$. However, at 
temperatures
below $T_U$, impurities are likely to be the dominant scatterers of 
electrons.  One can
 see this from the estimate
for the impurity contribution,  $\sigma_i \approx 5.5 \times 
10^{25}~(\rho_{14}
x/0.1)^{1/3}Z/(60~ Q)~ {\rm s^{-1}}$ (\cite{urp80}). Here $Q$ is the mean 
square
deviation of the atomic number from its average value.  The impurity 
parameter $Q$ is 
not well known, but the estimate of Flowers \& Ruderman (1977) suggests 
$\sim 10^{-3}$, to be a reasonable lower limit.

\section{Asymmetric accretion}

The problem of accretion onto neutron stars and the interaction with the
magnetosphere is a major research area in modern astrophysics and many 
interesting and
challenging problems remain to be solved, see for example Lamb (1991), 
Michel
(1991), White et al. (1995) for interesting discussions of this and 
related
phenomena.
I assume here the accreting matter to fall upon the neutron star in an
accretion column whose size is given by the last open field line before 
the
accretion disk. For simplicity, I also assume there be no ``slipping'' 
between
field lines as argued by Arons \& Lea (1980). The basic idea (Lamb et al. 
1973
and Shapiro \& Teukolsky 1983) is that matter is captured by some 
instability
mechanism at $r_A$ where the kinetic energy density of the accreting 
matter is
equal to the magnetic energy density, $r_A = (B_s^4 R^{12}/(2 G M \dot
M^2))^{1/7}\approx 3\times10^8 \dot M_{17}^{-2/7} B_{s,12}^{4/7} R_6^{12/7}
(M/M_\odot)^{-1/7}~{\rm cm}$, where $B_{s,12}$ is the surface magnetic 
field
strength in units of $10^{12}$ gauss, $\dot M_{17}$ is the accretion rate 
in
units of $10^{17}~{\rm g~s^{-1}}$, $R_6$ is the stellar radius in units 
of $10^6$
cm, and $M$ is the mass of the neutron star. Matter is then channeled by 
the
field lines onto the neutron star surface in a column with an opening
angle $\theta_{cap}$. The area under the accretion column is $A_{cap} = 
2\pi R^2
(1-\cos\theta_{cap})$. The angle $\theta_{cap}$ is given by those dipole 
field
lines that would, in the absence of accretion, have penetrated beyond 
$r_A$. 
The field lines for a dipole field are defined by $\sin^2\theta/r = {\rm
constant}$, so the base of the last undistorted field line which would 
close
inside $r_A$ lies at an angle 
\begin{equation}\label{eq:tetacap}
\sin^2\theta_{cap} = R/r_A = 3\times 10^{-3}\left[R_6^{-5/7} \dot M_{17}^{2/7}
B_{s,12}^{-4/7} (M/M_\odot)^{1/7}\right], \end{equation}
 for a detailed
derivation see e.g. Shapiro \& Teukolsky (1983). As the field strength 
decreases
due to ohmic losses the opening angle gets larger and larger. Eventually, 
the
field strength will be so low the accretion is expected to be spherical 
since
then $\theta_{cap}=\pi/2$. We have $B_{s,12}^{sph}=5\times 10^{-5}[
R_6^{-5/4}\dot M_{17}^{1/2} (M/M_{\sun})^{1/4}]$ where $B_{s,12}^{sph}$
is the value of the magnetic field when the accretion is spherical.

 This is of course a very simplified picture of the accretion
process, but it will nevertheless provide some insight as to how the 
crust
affects the surface magnetic field.

\subsection{Crude Temperature Structure} This section discusses some 
relevant
order of magnitude estimates. Although there are many heat sources in an
accreting neutron star we will use 
hydrogen burning in a thin shell at 
$\rho_{shell}\approx10^6~{\rm g cm^{-3}}$, see Mir90, as the major one.
Mir90 calculated models suppressing all other heat sources and came
up with temperature differences of a about a factor $2$. The density where the accreted
material starts to spread laterally, $\rho_{spread}$ is shown to be about 
an
order of magnitude higher than the base of the hydrogen burning shell,
$\rho_{shell}$. We also estimate typical thermal length scales to be 
shorter than
the crust thickness. These estimates indicate the heat source in our 
model to
have a size of about $l_{cap}\sim\sqrt{A_{cap}}$ and located at $\rho \approx 10^6~{\rm
g~cm^{-3}}$. Further, it is shown that the time for the temperature 
distribution to
reach a steady state is much shorter than typical magnetic field decay 
time
scales.

The kinetic energy of the accreting matter is turned into heat in the 
upper parts
of the atmosphere. After being 
thermalized
the matter continues to flow downwards at sub sonic speed. It will follow 
the
magnetic field lines down to a depth where the local magnetic force 
equals the
local pressure gradient. At this point the flow is presumed to spread 
laterally,
see Bildsten \& Brown (1996, 1997). They found $P_{spread}= B^2 l_{cap} /(4\pi 
h)$,
where $\l_{cap}\sim \sqrt{A_{cap}}$ and $h$ is a pressure scale height. 
If we
assume the pressure to be dominated by relativistic, degenerate electrons 
we find
$\rho_{spread} = 2~10^7~A_{cap,11}^{3/8}~B_{12}^{3/2} 
x_{0.5}^{-1}~h_4^{-3/4}
~{\rm g~cm^{-3}}$. As the accreted material flows down along the field 
lines its
density will increase which at certain thin shells will cause nuclear 
reactions
to occur which releases heat. For a thorough discussion of these matters 
see for example Brown \& Bildsten (1997), 
Bildsten et al. (1993), and Mir90. The base of the hydrogen burning shell
$\rho_{shell}\approx 10^6~{\rm g~cm^{-3}}$ for the spherical accretion 
case, see
Mir90. In our model, the size of the polar cap $l_{cap} \gg h$ so one can 
expect
$\rho_{shell}$ to be similar to the spherical calculation. We see for 
typical
values of the relevant parameters $\rho_{spread}>\rho_{shell}$. However, 
it is
also clear that as the field strength is reduced $\rho_{spread}$ is 
decreasing
and at some some value of $B_s$, $\rho_{spread}$ may be less than 
$\rho_{shell}$
and the flow might start to spread before hydrogen ignites. The resulting 
temperature 
and density structure for that case
must ultimately be answered by a thorough magneto-hydrodynamical 
calculation,
and in this paper I will use a simple parametrization.

Assuming the energy generation in the accretion column to be dominated by
hydrogen burning at the $\beta$-decay limited value of the ``hot'' CNO 
cycle we
get from Mir90 a luminosity $L_{nuc} = 10^{33} (Z/0.001)^2 \dot 
M_{17}~{\rm
erg~s^{-1}}$, where $Z$ is the heavy element content, or a flux $F_{nuc} 
=
L_{nuc}/A_{cap}$. To find a typical length scale for temperature 
variations,
which we denote by $l_{T}$, one can equate this flux with the thermal 
heat flux
$F_{heat} = -K \Delta T/l_{T}$, where 
$K=2~10^{18}\rho_{10}^{2/3}
(A/Z)^{2/3} ~{\rm erg~cm^{-1}~s^{-1}~K^{-1}}$ is the thermal conductivity
calculated by Urpin \& Yakovlev (1980) for the case where the $T>T_D$. The 
radiative contribution 
to the heat flux is negligible for the temperatures and densities of interest
here, Mir90 and M\'esz\'aros (1992). We 
get
$l_{T} = K \Delta T A_{cap}/ L_{nuc} = 10^2
\rho_{6}^{2/3} A_{cap,11} Z_{0.001}^{-2} \Delta T_8 \dot M_{17}^{-1}~{\rm 
cm}\;,$
 where I have normalized the density to $10^6~{\rm g cm^{-3}}$
which is roughly the value of $\rho_{shell}$. $A_{cap,11}$ is the area
of the accretion cap in units of $10^{11}~{\rm cm^2}$ and 
$Z_{0.001}=Z/0.001$ . 
We see that $l_T$ is typically smaller than a crust thickness, $\Delta 
R\approx
10^5{~\rm cm}$ and for some parameter values it can be much smaller.
When $l_T\ll\Delta R$ or $l_T\gg\Delta R$ the temperature structure in the 
crust is
spherical which has been studied elsewhere, see Mir90. Here, I will focus
on the intermediate case, where $l_T \leq {\rm a~few~}\Delta R$.

To get an estimate of the temperature
structure below and around the accretion cap we can simply
note that as the conductivity increases inwards the thermal lengthscale also increases
so for a given temperature change,  $\Delta T$, we need to go a distance 
$l_{T,r} \sim \bar K\Delta T$  
radially which
is greater than the corresponding horizontal distance 
$l_{T,\theta}\sim \Delta T K$ where $\bar K=(\int_r^{r+l_{T,r}} K^{-1} dr)^{-1}
l_{T,r}$.
In other words, the
hydrogen
burning shell will produce a temperature structure that has an ellipsoidal
or, as I will assume in this paper, columnar 
shape. 
In the estimate of $l_T$ we have simply assumed
the relevant area to be $A_{cap}$ for all densities, and so ignored
the heat flowing through the sides of the column. However, since the size of the
polar cap is of the same order as the crust thickness this correction
is likely to be small.
Given the uncertainty of the horizontal lengthscale and the size
of the hot accretion cap I choose to
parameterize $l_T$ in this paper, see section 4.

The time scale for reaching a steady state is now given by, $\tau_T = c_p
h^2\rho/ K=8~10^7 \rho_{10}^{1/3} x_{0.5}^{-2/3} l_{T,4}^2/A~{\rm s}$ 
where
we have assumed $T_{cap}>T_D$ which holds below $\rho\approx 
3~10^9~T_{cap,8}^2
~x_{0.5}^{-2} ~{\rm g~cm^{-3}}$. $c_p$ is the specific heat capacity of 
the
crust, see e.g. Shapiro \& Teukolsky (1983) and $A$ is the atomic number. 
The
magnetic field decay time is given by $\tau_{B}=4\pi \sigma l_T^2/c^2$ 
which with
$\sigma=\sigma_i$ is $\tau_{B,i} = 5.5 \times 10^{17}~(\delta 
R_5)^2~(\rho_{14}
x/0.1)^{1/3}Z/(60~ (Q/0.01))~{\rm s}$. When the accretion is spherical
$\sigma=\sigma_{ph}$ and $\tau_{B,ph}=5.5 \times 10^{15}~ (\delta
R_5)^2~\rho_{14}^{7/6} ~(x/0.1)^{5/3}T_8^{-2}~ {\rm s}$. We see that
$\tau_T\ll\tau_B$ when the conductivity is dominated by either phonon or 
impurity
scattering. In the estimates of the field decay times we have normalized 
the
values of the parameters to those of the inner crust. The temperature 
will thus
reach a steady state long before the magnetic field has changed 
significantly.

Given these estimates it is likely that the temperature structure in the 
crust of
a neutron star, accreting in a column, is such that inside the accretion 
column
the temperature resembles the calculation by Mir90 and by simply 
extrapolating
their result we find, 
\begin{equation}\label{eq:Tcap} T_{cap}\approx
10^8~\left({\dot M\over{10^{17}{\rm ~g ~s^{-1}}}}{4\pi R^2\over
A_{cap}}\right)^{1/4}~ {\rm K} \;. 
\end{equation} 
Outside the column the
temperature is much lower. Precisely how much lower remains to be seen by
detailed calculations but motivated by the previous estimates of $l_T$
we will in this paper assume it to be low 
enough for
impurity scattering to be the dominant source of electrical conductivity
$\sigma=\sigma_i$. 

I use for the conductivity, 
\begin{equation}\label{eq:conductivity}
 \sigma(\theta) =
{\sigma_i-\sigma_{cap}\over {1+
\exp\{(C_{orr}+|\theta-\pi/2|-(\pi/2-\theta_{cap}))/ \beta\} } } 
+\sigma_{cap} ,
\end{equation} 
where $\sigma_{cap}=\sigma_{Um}~,~T<T_D$, $\sigma_{cap} =
\sigma_{ph}~,~ T>T_D$ and $C_{orr}=(\theta_{cap}/(\pi/2))^{20}$ is a 
correction
used to ensure that $\sigma_\theta=\sigma_{cap}$ when the accretion is 
uniform
over the whole sphere $\theta_{cap}=\pi/2$.
To see how the 
size of the
horizontal thermal length scale affect the solution I use a simple
parameterization of $l_T=R \beta$ where $\beta$ is assumed to be 
independent of
$r$ and in this paper range from $0.02 - 0.2$. 

It is worth pointing out that if the different parameters 
are such that
$l_T \ll \Delta R$ initially, $l_T$ will increase as the 
magnetic polar cap
size increases and could eventually be such the whole crust is heated up.
The spatial evolution of the magnetic field in the crust will then be
different from this calculation. However, the {\it temporal} evolution,
will be very similar, that is the timescale for the accretion to
be completely spherical is the same as we find here, and also
the subsequent rapid field decay. 

\section{Numerical Results}

This section describes numerical solutions to equations (\ref{eq:Psi1}) 
and
(\ref{eq:conductivity}) with two
different initial conditions. 
First, I choose a configuration where the field is penetrating the 
whole crust,
 $\rho\leq 2.4~10^{14}~{\rm g~ cm^{-3}}$.
 In the second case the field was initially confined to $\rho\leq 
10^{11}~{\rm g~
cm^{-3}}$. In both cases the initial surface dipole strength was $10^{12}$ gauss. 

The results are combined into three sections where in the first, section 4.1, the
field strength is initially penetrating the whole crust. This is followed
by section 4.2 where the field fills only the outer crust $\rho\leq 10^{11}~{\rm
g~cm^{-3}}$ initially.
 The numerical
calculations show the decay time scale can be divided into two stages where
in stage I the decay time scale is
given by $\tau_{B,i}$ and when the accretion becomes spherical it enters 
the second stage where the decay rate
increases to $\tau_{B,ph}$. The calculations also show that starting from 
a dipole,
the contribution from higher order surface multipoles is at the most 
$30~\%$ 
to the dipole mode in agreement with Arons (1993). In section 4.3 I explore different values of the horizontal
length scale by varying $\beta$.

Equations (\ref{eq:Psi1}) and (\ref{eq:conductivity}) was solved with the 
finite-difference code described
elsewhere, Sahrling (1996). For the impurity parameter I chose four 
values
$Q=0.001,0.01,0.1,$ and $1$ which span the likely range of values, see 
Flowers \&
Ruderman (1977), and is constant throughout the crust. The model I am 
considering
here is an $M= 1.4 M_\odot$ neutron star with the equation of state given 
by
Wiringa et al. (1988) and Baym et al. (1971), with radius $R=11.5 {\rm 
km}$. The equation of state for neutron stars is a debated issue but this choice 
represents an average stiffness of the range of models discussed. 

At $t=0$ the field was set to be dipolar, $A_\varphi\sim \sin\theta$. Since
the conductivity below the accretion cap $\theta\leq\theta_{cap}$ and 
$\theta\geq \pi - \theta_{cap}$
is smaller than the conductivity outside, the field will initially change
more rapidly there producing higher-order multipoles.

Generally, it was found the dependence of the various runs on the 
impurity
parameter $Q$ is simple. Let us denote by $\tau_Q$ the time scale for 
runs with a
particular $Q$. Then $\tau_Q\approx\tau_{Q=1}/Q$ which given our 
approximations
is to be expected. From the observational constraints discussed 
in the introduction 
we can infer timescales for accretion lies roughly between 
$10^7-10^9~{\rm yrs}$ depending on such things as mass of the companion
star and the spin period $P_i$. 
The numerical results of our model indicates that the value of $Q$ has to
be greater than about 0.1 to give a significant decay of the magnetic field.

\subsection{Field initially penetrating the whole crust} 
The field strength was chosen to be $10^{12}$ gauss at the
stellar surface initially. This high field is consistent with the pattern 
of
High-Mass X-ray binaries (HMXBs), in which the X-rays originate in a 
small steady
hot patch, see Lyne \& Graham-Smith (1990). There is however one
LMBP system known, PSR 0820+02, with a similar high field strength, so
 this initial condition can also be relevant for LMXB's.
 To calculate $\theta_{cap}$ 
from
equation (\ref{eq:tetacap}) we use a constant $\dot M = 10^{17} ~{\rm 
g~s^{-1}}
\approx 10^{-9} M_\odot {\rm yr^{-1}}$. 
The parameter $\beta=0.02$ for 
all runs
discussed here.

Figure 1 shows the evolution of the various surface magnetic multipoles,
normalized to the initial value $B_s$. It shows the evolution for a crust
with $Q=1$. This high impurity level results in very low conductivity and 
thus
the most rapid field decay among the cases considered. Changing the value 
of $Q$
resulted only in a change in time scale of a factor $1/Q$. As can be seen 
from
the figure high-order multipoles are initially stronger than multipoles
$l=3, 5$ but
as time progresses they disappear more and more and the lower ones start 
to
dominate. The contribution of $l=3, 5$ compared to the dipole is about 10 
\% at
the time the accretion is spherical,
which happens when $B_s\approx 10^8~{\rm gauss}$ and the field evolution enters stage II. We saw earlier that
in our model this value varies with the square root of the accretion rate.
One has to conclude that if this mechanism is responsible for there
being no pulsars with a magnetic field lower than $10^8$ gauss the 
accretion rate for the majority of the millisecond pulsars during their recycling phase will have to have been,
on average,
greater than roughly $10^{17}~{\rm g~ s^{-1}}$. Observationally it is hard to pin point
the precise accretion rate but assuming the luminosity to be due solely
to the infalling matter one finds accretion rates ranging from $10^{16}-
10^{17} ~{\rm g ~s^{-1}}$ Phinney \& Kulkarni (1994).

\subsection{Field initially buried in outer crust}

If one instead starts out with a field buried below $\rho\leq 
10^{11}~{\rm g~
cm^{-3}}$ one finds the result presented in figure 2. There is an 
initially rapid
decay because the field diffuses into the interior of the star on a time 
scale,
$\tau_{B,i}$ calculated at $\rho=10^{11}~{\rm g~ cm^{-3}}$. The field 
starts to
penetrate the whole crust at a time scale given by the conductivity of 
the inner 
crust. This
is all what can be expected from dimensional analysis of the induction 
equation.
The $l=3, 5$ multipole strengths are in this case also about 10 \% of the dipole 
when
the accretion is spherical.

\subsection{The dependence on the horizontal thermal length scale}

To investigate the sensitivity of the solution to different horizontal 
thermal
length scales we used three different values of the parameter $\beta = 
0.02, 0.1,
0.2$. To avoid numerical problems I chose not to go below $\beta = 0.02$. 
The
initial field was assumed to penetrate the whole crust and have an 
initial
surface strength of $10^{12}$ gauss. The results are shown in figure 3 
a),b). These
plots should be compared to Fig. 1. As can be expected, the higher order
multipoles are less significant when $\beta$ increases. Other than that 
the
overall evolution is very similar for all values of $\beta$ considered.

\section{Conclusion} We have examined the influence of asymmetric 
accretion on
the magnetic field evolution of a neutron star. The temperature structure 
in the
crust resulting from the accretion was roughly estimated and found to 
vary more
rapidly than the crust thickness around the edge of the accretion cap. 
The time
scale for reaching a steady state was shown to be shorter than the time 
scale for
magnetic field decay. Therefore, instead of doing a full blown 
calculation of
heat conduction coupled with the magnetic field evolution we used a 
simple
smoothed top-hat function for the temperature structure and consequently 
for the
conductivity. The global field decay time is roughly the one given by the 
largest
conductivity in the crust, which occurs outside the accretion cap, 
$\tau_{B,i} =
5.5 \times 10^{17}~(\delta R_5)^2~(\rho_{14} x/0.1)^{1/3}Z/(60~ 
(Q/0.01))~{\rm
s}$. However, when the field strength is down to roughly $10^8$ gauss the
accretion is spherically symmetric and we have $\tau_{B,ph} = 5.5 \times 
10^{15}~
(\delta R_5)^2~\rho_{14}^{7/6} ~(x/0.1)^{5/3}T_8^{-2}~ {\rm s} \ll 
\tau_{B,i}$
and so the decay rate is much faster than initially which could provide a 
clue
why no binary pulsars are known with field strengths less than $10^8$ 
gauss. Romani (1993) discussed a similar effect and found a threshold
close to ours.
 We also found the asymmetric accretion
resulted in some higher order surface magnetic multipoles and the size of 
these
were shown to be at most 30 \% of the surface dipole field.

{\bf Acknowledgements :} I would like to thank Dong Lai, Lars Bildsten and 
Edward Brown for useful
discussions. Dong Lai is especially thanked for a careful reading of the
manuscript. This work was also supported in part by the U. S. National 
Science
Foundation under grants NSF AST93-15133 and AST94-14232, by NASA under 
grant
NAGW-1583, and by the Swedish Natural Science Research Council.

\vfill\eject

\vfill\eject\onecolumn

\centerline {\bf FIGURE CAPTIONS} \vskip1cm

\indent{\bf Figure 1 :}
The evolution of the various surface magnetic multipoles,
normalized to the initial value $B_s=10^{12}~{\rm gauss}$ with $Q=1$ and 
$\beta=0.02$. Initially
the filed penetrated the whole crust.
High-order multipoles are 
initially stronger than low-order 
ones but
as time progresses they disappear more and more and the lower ones start 
to
dominate. The dips in the curves corresponds to times when the multipoles 
changes sign. The contribution of $l=3, 5$ compared to the dipole is about 10 
\% at
the time when the accretion is spherical. The early part of the evolution from
$t=0$ is not shown here since that describes the initial transients of the
field due to the choice of initial conditions $A_\varphi \sim \sin\theta$.

\vskip1cm
\indent{\bf Figure 2 :} Same as figure 1 but with the initial field confined to $\rho\leq
10^{11}~{\rm g~cm^{-3}}$.

\vskip1cm
\indent{\bf Figure 3: } a) Same as figure 1 but with  $\beta=0.1$ initially. b)
Same as figure 1 but with  $\beta=0.2$ initially.

\vskip1cm

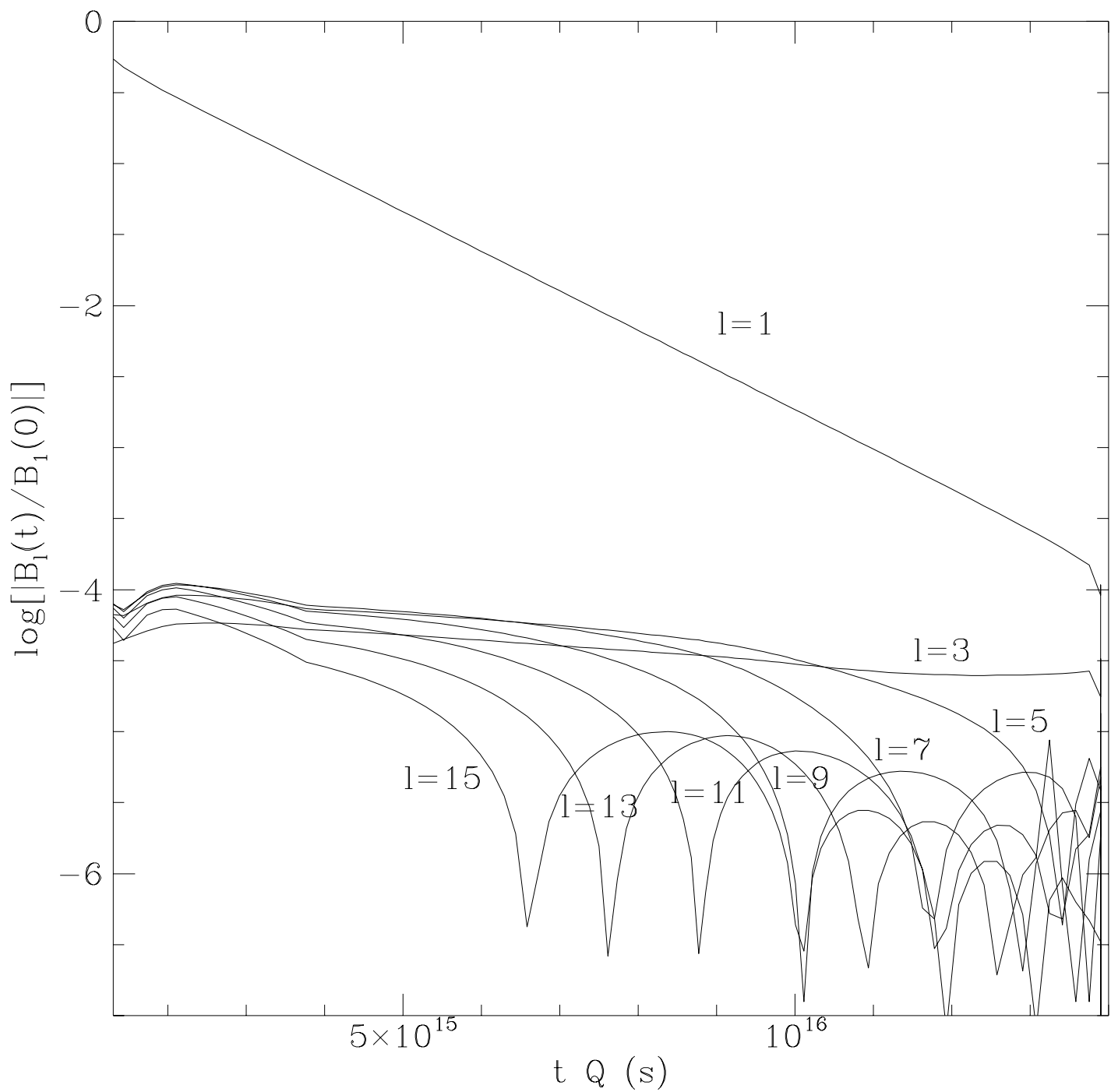
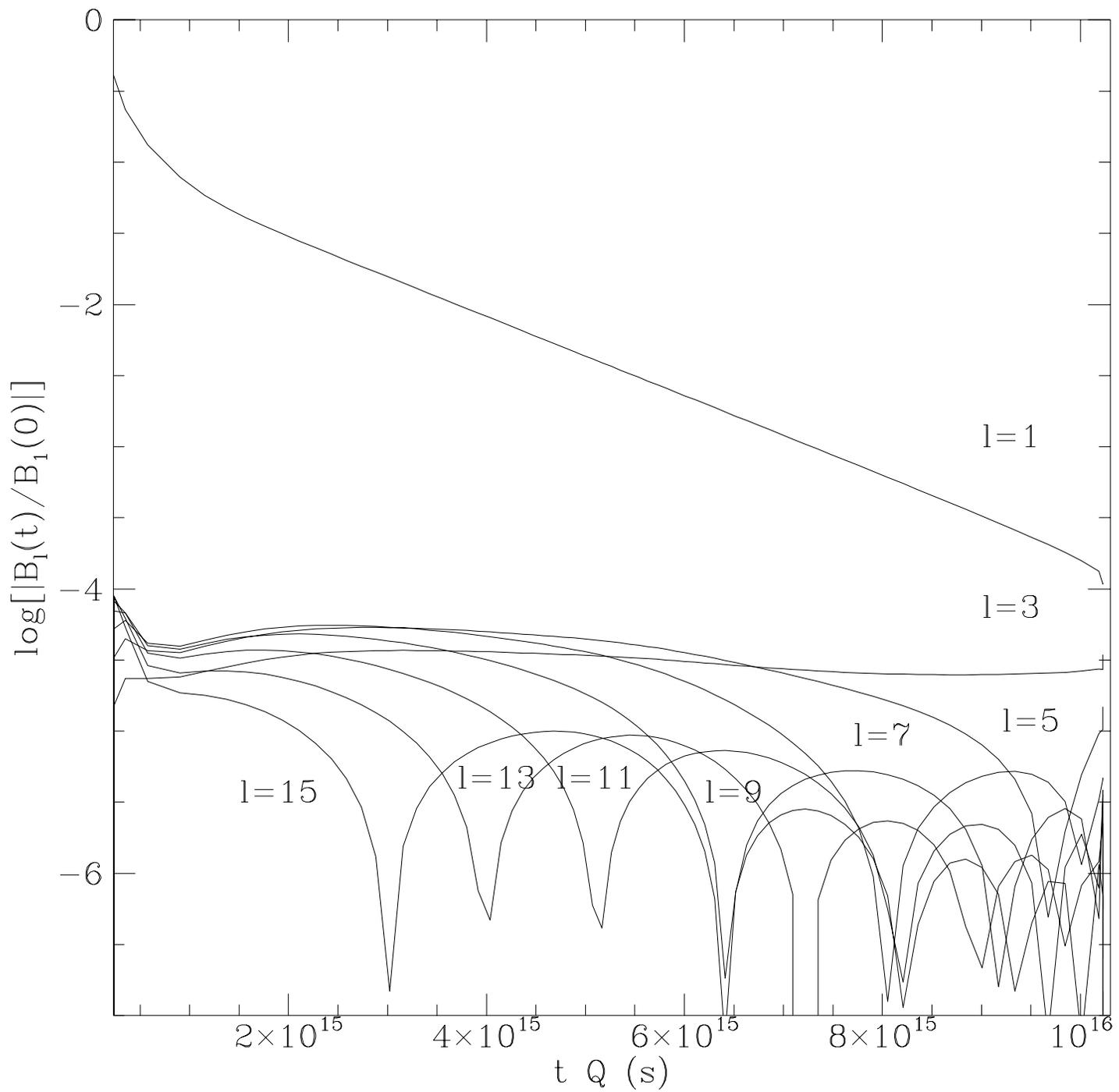
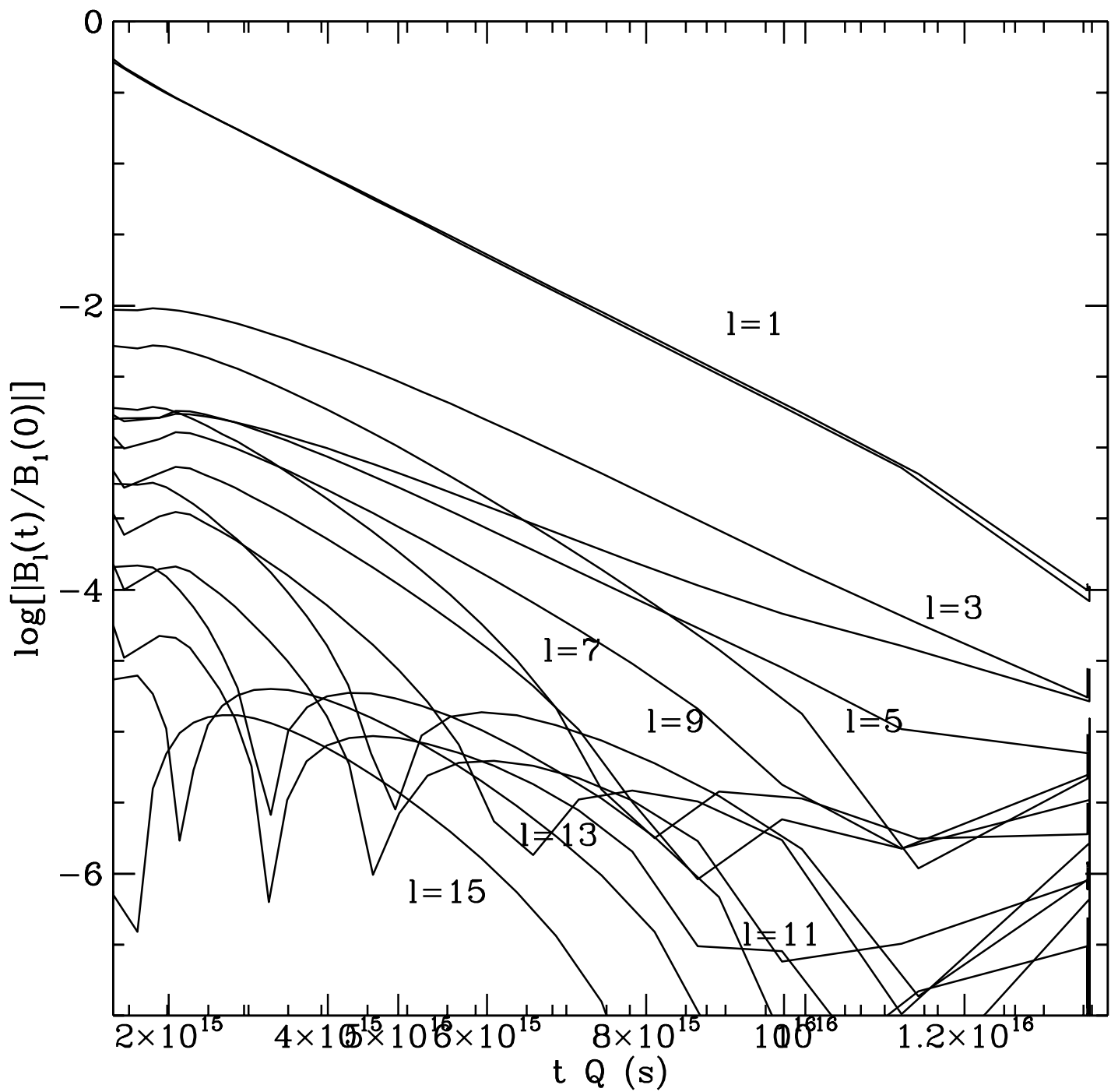
\end{document}